\documentstyle[12pt]{article}

\begin{document}
\thispagestyle{empty}
\begin{center}
\LARGE \tt \bf{Placing limits on the spin-torsion fluctuation from the stability of Friedmann solution and COBE data}
\end{center}
\vspace{1cm}
\begin{center}{\large L.C. Garcia de Andrade\footnote{Departamento de
F\'{\i}sica Te\'{o}rica - Instituto de F\'{\i}sica - UERJ
Rua S\~{a}o Fco. Xavier 524, Rio de Janeiro, RJ
Maracan\~{a}, CEP:20550-003 , Brasil.
E-Mail.: GARCIA@SYMBCOMP.UERJ.BR}}
\end{center}
\vspace{1.0cm}
\begin{abstract}
The gravitational stability of Friedmann metric with respect to homogeneous and isotropic perturbations in Einstein-Cartan cosmology is used together with the COBE data to place a limit on the spin-torsion primordial density fluctuation.Cosmological perturbations in Bianchi type III models are shown to have no contributions from torsion at the early stages of the Universe.
\end{abstract}      
\vspace{1.0cm}       
\begin{center}
\Large{PACS number(s):04.20.Dw}
\end{center}
\newpage
Ealier Nurgaliev and Ponomariev \cite{1} investigated the early evolutionary stages of the Universe in the realm of Einstein-Cartan cosmology by considering an ideal fluid of nonpolarised fermions.They show that for this particular case the Friedmann solution was stable to small homogeneous and isotropic perturbations.They also conclude that the increase in entropy could lead to the evolution of the initially small stable oscillations into large ones.The Universe would begin after a definite number of oscillations.Calculations of small perturbations in their model would shown some instability which would prepare for the necessary initial conditions for the growth of pertubations in the nonlinear regime.In this note we make use of their result to olace a lower limit to the spin-torsion primordial density fluctuation obtained from the Einstein-Cartan gravity and COBE satellite data \cite{2,3}.In this way structure formation like Galaxies formation would not suffer a great influence from torsion \cite{3}.It is also shown that at early stages of the Universe Bianchi type III models with expansion and rotation may not depend at all from torsion \cite{4}.Other Bianchi types like an oscillating Bianchi type IX model in Einstein-Cartan \cite{5} gravity have also been recently investigated.
\begin{equation}
ds^{2}=dt^{2}-R^{2}(t)(dz^{2}+dx^{2}+dy^{2})
\label{1}
\end{equation}
The Friedmann equation thus becomes
\begin{equation}
\frac{{\ddot{R}}}{{R}}=\frac{4{\pi}G}{3}({\rho}-8{\pi}G{\sigma}^{2})
\label{2}
\end{equation}
Making an homogeneous and isotropic small perturbation on the Friedmann yields
\begin{equation}
\frac{{\delta}{\ddot{R}}}{{\delta}{R}}=-\frac{4{\pi}G}{3}({\rho}-8{\pi}G{\sigma}^{2})-\frac{4{\pi}G}{3}(\frac{{\delta}{\rho}}{{\delta}R}-8{\pi}G\frac{{\delta}{\sigma}^{2}}{{\delta}{R}})
\label{3}
\end{equation}
Substitution of the well-known relation
\begin{equation}
\frac{{\delta}{\rho}}{{\delta}R}=-3\frac{\rho}{R}
\label{4}
\end{equation}
we obtain
\begin{equation}
\frac{{\delta}{\ddot{R}}}{{\delta}{R}}=-\frac{4{\pi}G}{3}({\rho}-8{\pi}G{\sigma}^{2})+\frac{4{\pi}G}{3}(\frac{{\rho}}{R}+8{\pi}G\frac{{\delta}{\sigma}^{2}}{{\delta}{R}})
\label{5}
\end{equation}
and the stability condition $\frac{{\delta}{\ddot{R}}}{{\delta}{R}}<0$ implies 
\begin{equation}
(1-8{\pi}G\frac{{\delta}{\sigma}^{2}}{{\delta}{\rho}})<0
\label{6}  
\end{equation}
Since the matter density ${\rho}>0$ this implies the following condition
\begin{equation}
{{\delta}{\sigma}^{2}}>\frac{1}{8{\pi}G}\frac{{\delta}{\rho}}{{\rho}_{0}}{\rho}_{0}
\label{7}  
\end{equation}
where we take${\rho}_{0}=10^{-30}g$ as the matter density of the Universe and from the COBE data $\frac{{\delta}{\rho}}{{\rho}_{0}}=10^{-5}$.From these data formula (\ref{7}) yields the following lower limit for the spin-torsion fluctuation as
\begin{equation}
{\delta}{\sigma}^{2}>10^{-27} cgs units
\label{8}
\end{equation}
 this result was expected since the spin-torsion density decreases with the expansion and is redshifted with inflation.This conjecture has been proposed recently by Ramos and myself \cite{6}.As pointed out by Nurgaliev and Ponomariev \cite{1} the increase in the entropy may trigger the growth in the inhomogeneities.There is no compelling reason to believe that this would not happen here.Now let us note that in the case of the Bianchi III of a rotating and expanding Universe \cite{7} 
\begin{equation}
ds^{2}=dt^{2}-2R{{\sigma}_{0}}^{\frac{1}{2}}adydt-R^{2}(dx^{2}+ka^{2}dy^{2}+dz^{2})
\label{9}
\end{equation}
where R(t) is the cosmic scale;$a=exp{mx}$;$m>0$,$k>0$ and ${\sigma}_{0}$ are parameters.This metric in fact poossess expansion,shear and rotation and gives a much more complete description of the stages of the Universe.The Einstein-Cartan field equations of gravity yields
\begin{equation}
\frac{{\dot{R}}^{2}}{R^{2}}=\frac{{\alpha}_{1}}{R^{2}}+\frac{{\alpha}_{2}}{R^{6}}+{{\alpha}_{3}}+\frac{{\alpha}_{4}}{R^{4}}
\label{10}
\end{equation}
and the energy density is
\begin{equation}
{\rho}=-\frac{{\Lambda}}{k}-\frac{A}{4k}+\frac{3}{2}\frac{B^{2}}{R^{4}}-\frac{S^{2}_{0}}{8a_{1}R^{6}}
\label{11}
\end{equation}
where $S_{0}^{2}$ is the spin-torsion density at the beginning of the Early Universe.Computation of the cosmological perturbation density reads 
\begin{equation}
{\delta}=\frac{{\delta}{\rho}}{\rho}=1-\frac{{R'}^{6}}{R^{6}}
\label{12}
\end{equation}
and as one notices there is no dependence of Cartan torsion or spin.Of course this is not a general result valid in all stages of the Universe.
\begin{equation}
{\rho}'=-\frac{{\Lambda}}{k}-\frac{A}{4k}+\frac{3}{2}\frac{B^{2}}{{R'}^{4}}-\frac{S^{2}_{0}}{8a_{1}{R'}^{6}}
\label{13}
\end{equation}
The gravitational instability of de Sitter metric has been proved in the case of higher dimensional gravity with torsion by Maroto and Shapiro \cite{8}.Moreover Nurgaliev and Piskareva \cite{9} have also investigate the structural stability of cosmological models in Einstein-Cartan gravity.A more detailed investigation of the matters discussed here including the Bianchi type IX oscilating solution in Einstein-Cartan cosmology may appear elsewhere.
\section*{Acknowledgments}
I am very much indebt to Y.Obukhov,R.Ramos,P.S.Letelier and I.Shapiro for helpful discussions on the subject of this paper.Financial support from Universidade do Estado do Rio de Janeiro (UERJ) and CNPq is grateful acknowledged.

\end{document}